\documentclass[12,dvips,twocolumn]{article}
\usepackage[latin1]{inputenc}
\usepackage{xspace}
\usepackage{graphicx}
\usepackage{xspace}
\def\htsc{high temperature superconductors\xspace}
\def\Dx{\ensuremath{\Delta x}\xspace}
\def\Dt{\ensuremath{\Delta t}\xspace}
\newcommand{\T}[2]{{\ensuremath{T_{#1} ^{#2}}}}
\def\pui{\textsuperscript}

\begin{document}
\title{
  \fbox{
    \begin{tabular}{c}
      Calorimetric study of thermal properties \\
      of superconducting tapes~:\\
      experimental method and simulations.
    \end{tabular}
    }
}

\author{\begin{tabular}{l}
Benoît des Ligneris$^a$, Marcel Aubin$^a$, Julian Cave$^b$\vspace{12pt}\\
$^a$ {\footnotesize Département de physique and Centre de recherche sur les propriétés électroniques de matériaux avancés}\\
{\footnotesize Université de Sherbrooke, Sherbrooke, Québec, Canada J1K 2R1}\\
$^b$ {\footnotesize Technologies Émergentes de production et de stockage, VPTI Hydro-Québec}\\
{\footnotesize Québec, Canada J3X 1S1}
\end{tabular}
}

\date{
\begin{tabular}{l}
June 13 2000\\
To be published in {\em Proceedings of the ICMC2000, Rio de Janeiro 11-15 june 2000}\\
\\
\begin{tabular}{ll}
\fbox{Keywords~:} & calorimetric measurement, thermal conductivity, \\
&  high temperature superconductivity, lumped capacitance, finite element
\end{tabular}
\end{tabular}
}

\maketitle{}

%\begin{center}
%\begin{minipage}{5in}
 
%\fbox{\huge\begin{tabular}{c}
%Calorimetric study of thermal properties \\
%of superconducting tapes~:\\
% experimental method and simulations.
%\end{tabular}}
%\end{minipage}
%\end{center}

%\vspace{2cm}

%\begin{minipage}{6in}
%{Benoît des Ligneris$^a$, Marcel Aubin$^a$, Julian Cave$^b$\vspace{12pt}\\
%{$^a$ \footnotesize Département de physique and Centre de recherche sur les propriétés électroniques de matériaux avancés}\\
%{ \footnotesize Université de Sherbrooke, Sherbrooke, Québec, Canada J1K 2R1}\\
%{ $^b$ \footnotesize Technologies Émergentes de production et de stockage, VPTI Hydro-Québec}\\
%{ \footnotesize Québec, Canada J3X 1S1}}

%\end{minipage}

%\vspace{1cm}

\begin{abstract}

The method consists in the monitoring of the temperature variation of a superconductor tape subjected to a trapezoidal current pulse. Simulations of the experiment were performed using the lumped capacitance model and also the finite element method. The former reproduces the sample's reaction for the increasing and decreasing parts of the pulse. The same can be said for the finite element method but the decreasing temperature as expected is not adequately reproduced.  Treating the problem in two dimensions rather than one should correct the situation. Nethertheless we obtained a thermal conductivity consistent with the literature.

\end{abstract}

\section{\sc\bf Introduction}

Thermal properties are a very good indicator of the superconductive transition. However, from a practical point of view, the variation of the operating temperature is usually very small (say around 77.4~$\pm$~0.5~K)  for power applications. Thus the thermal properties of \htsc are usually supposed constant when the system is under normal operating conditions. Actually, the main sources of temperature variation are the losses which can be separated into a hysteretic contribution (AC losses) and other losses that occur because of thermally activated phenomena, losses in the normal matrix, ... The true losses are the complete losses (sum of those two contributions) and cannot be measured by conventional electrical measurements. In order to reproduce such a phenomenon without knowing the explicit dependence of the true losses with respect to external parameters (applied field (DC and AC), applied current (DC and AC), temperature, ...), we will isolate a part of the sample and then generate heat internally. We will then try to modelize the system in order to determine to what degree the thermal properties are affected.

\section{\sc\bf Description of the method}

\begin{figure}[pt]
  \begin{center}
    \includegraphics[width=3in]{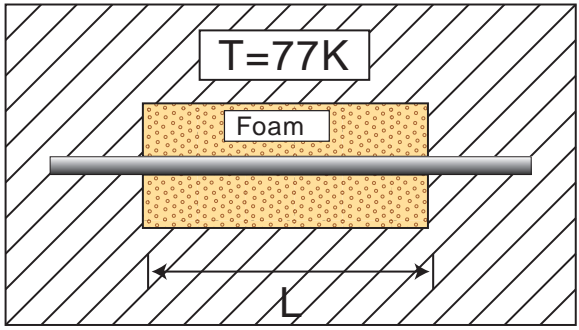}
    \caption{\footnotesize  Principle of the experiment}
    \label{fig:principe:experience}
  \end{center}
\end{figure}

The experiment is very simple and its principle is represented in Fig.\ref{fig:principe:experience}.

Part of the BSCCO tape is sandwiched between the two halves of a polystyrene block. The joints are covered with a silicon adhesive so that no liquid nitrogen will reach the protected part of the tape when the block is immersed in the cryogenic bath.

The data reported here were obtained with a BSCCO tape from BICC. A CERNOX sensor from Lakeshore Cryotronics was placed at the tape center to measure the temperature inside the block. We believe that the higher complexity of the experiment using a CERNOX sensor instead of a thermocouple is compensated by the absolute temperature measurement read by the CERNOX sensor (compared to relative measurements for thermocouples).

The temperature of the nitrogen bath is not constant for two major reasons~: the stratification of the fluid can change the local temperature and the liquefaction of ambient oxygen (with a boiling temperature of 90K) can change the average temperature of the bath. This drift cannot be evaluated with only one sensor~: ideally, one would include another sensor outside the block (soldered to the sample) in order to monitor the thermal drift.

In this article, data from only one sensor are used but we are currently using three sensors distributed along the protected part of sample. Sensors are mounted on small sapphire rectangles placed transversely across the tape so that we take into account all the heat contributions from a given cross-section. The thermal drift was substracted from the data using linear interpolation between the initial and final temperatures.

\begin{figure}[pt]
  \begin{center}
    \includegraphics[height=3 in,angle=270]{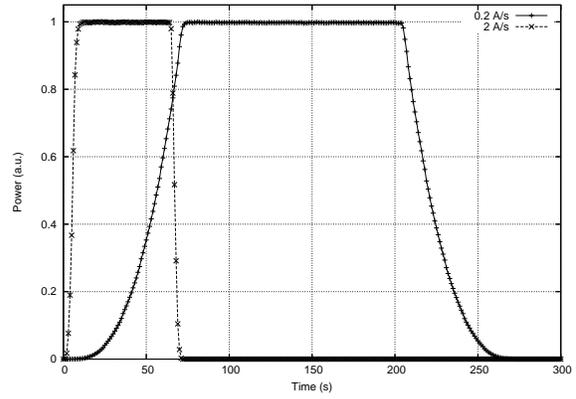}
    \caption{\footnotesize Power pulses with current ramps of 2A/s and 0.2A/s}
    \label{fig:current:pulse}
  \end{center}
\end{figure}

Our experimental data consist of two measurements, both obtained from the same sample, in the same cryostat and with the same maximum current (13 A). The shape of the current pulse applied is trapezoidal. In one case the slope is 2A/s and in the other 0.2A/s. The resulting power pulses are represented in Fig~\ref{fig:current:pulse}.

Below, we shall compare our result for the thermal conductivity with those of ref~\cite{thermique1} and \cite{goldacker2000}. These depends strongly on the type of sheath that envelops the superconductor.

The specific heat varies from sample to sample but, according to \cite{castro2000}, we have, for typical BSCCO tapes, $0.150<C_p (J.g^{-1}.K^{-1})<0.190$. A 10\%  margin is to be considered for the same sample because of eventual inhomogeneities.

\section{\sc\bf Numerical Calculations}

In this section, we will address the problem by two different methods~: firstly the so called ``lumped capacitance model'' and then a more numerical approach~: the explicit 1D finite element method.

\subsection{The lumped capacitance model}

For a review of heat transfer, see ref~\cite{fundamentals:heat:mass}. The main hypothesis of this model is that the temperature is spatially uniform. This implies that temperature gradients within the solid (in this case the superconductor) are negligible. Thus we cannot use the standard framework of heat conduction expressed by the Fourier law (Eq.~\ref{eq:loi:de:fourier}). Instead, we calculate the {\em transient} response of the temperature T which is determined by formulating the overall energy balance of the solid. So if the solid is in a fluid (namely nitrogen) with a convection factor $h\ (W.m^{-2}.K^{-1})$ at a temperature $T_\infty$, then the equation $E_{storage}=-E_{out}$ becomes~:
\\

\begin{equation}
\rho V C_p \frac{dT}
                {dt}\ =\ -h A_s (T - T_\infty )
\end{equation}

\noindent $A_s$ is the surface of the sample in contact with the convective fluid, $\rho$ $( kg.m^{-3} )$, $V$ $( m^3 )$, $C_p$ $( J.kg^{-1}.K^{-1} )$  are the density, the volume and the specific heat of the sample respectively and $t$ ($s$) the time.

In this case, we can define a thermal time constant for the system~:

\begin{equation}
\tau_t = \left( \frac{1}
                     {hA_s}\right)(\rho V) = R_t C_t
\end{equation}

\noindent where $R_t=\left( \frac{1}
                                 {hA_s} \right)$ is the resistance to convection and $C_t=\rho V$ is the lumped thermal capacitance.

In order to know if such a regime is reached, we must examine the dimensionless Biot number, $Bi=(h L_c)/(k)$ where $L_c$ is the characteristic length of the problem and $k$ is the thermal conductivity $(W.m^{-1}.K^{-1})$. In order to apply the lumped capacitance model, the Biot number must be very small compared to unity. If this is the case, we can say that the resistance to conduction within the solid is much less than the resistance to convection across the fluid boundary layer.

In our case, the studied system is more complex but if the same representation can be used, we can modelize our system by the sample, the sample holder and the global environment ($h, T_\infty$). If there are heat leaks in the sample holder this will lead to two time constants rather than one~: one time constant is due to the unidimensional conduction (exchange of heat at the extremities of the sample) and the other one due to the heat exchange between the sample and the sample-holder. 

The net outcome of this method is an exponential behavior. By fitting the data with exponentials we implicitly assume that this model is correct and we modelize the sample holder action by introducing the second time constant. Even if the most interesting information is surely the sample's contribution, that of the sample holder must be studied if we want to ensure that our sample is ``well'' isolated. Moreover, we will be able to determine a characteristic time constant of our system (sample + sample holder). After this time, our system can no longer be considered isolated since the heat flow through the sample holder must be taken into account.

Another aspect of our problem is that the heat is generated internally by an electric current and then flows outside the system. If the heat generated is important (compared to the leaks) and the heating-time is short (which means that the heat exchange between sample and sample-holder is small compared to internally generated heat) we can neglect the second time constant for the heating period only.

\begin{figure}[pt]
  \begin{center}
    \includegraphics[width=3 in]{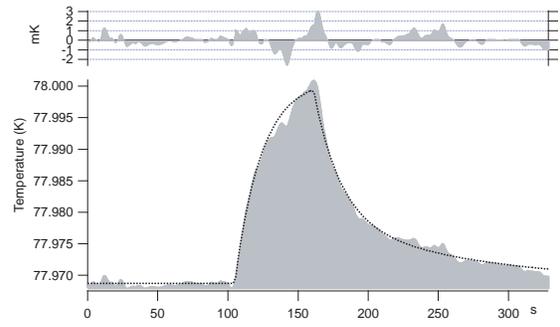}
    \caption{\footnotesize  Fit with two exponentials (lumped capacitance model), following the application of the 2A/s trapezoidal current pulse. The upper part represents the error between the data and the fit}
    \label{fig:fit:deux:exp:1}
  \end{center}
\end{figure}

The fit with two exponentials, for the 2A/s case is represented in Fig.~\ref{fig:fit:deux:exp:1}. Similar results were obtained with the 0.2A/s pulse.
 
\subsection{1D explicit finite element modelization}

For a review of the finite element method, see ref~\cite{elements:finis}. The master equation is Fourier's law~:

\begin{equation}
k A_s \frac{dT}{dx} + \dot{q} V = \rho V C_p \frac{dT}{dt}
\label{eq:loi:de:fourier}
\end{equation}

In order to use the finite element method, the space is ``discretized'' in spatial and time intervals, $\Delta x$ and $\Delta t$. Then we can write~:

\begin{equation}
\begin{array}{c}
k A_s \frac{\T{m-1}{p}-\T{m}{p}}
           {\Dx} + 
k A_s \frac{\T{m+1}{p}-\T{m}{p}}
           {\Dx} +
\dot{q} A_s \Dx = \\
\rho A_s \Dx C_p \frac{\T{m}{p+1}-\T{m}{p}}{\Dt} 
\end{array}
\end{equation}

Isolating the (\T{m}{p+1}) term, we can write the following equations for all the nodes except the first and last ones~:

\begin{equation}
\begin{array}{c}
\frac{k}
     {\rho C_p (\Dx)^2} 
\left( \T{m-1}{p}+\T{m+1}{p}+\frac{\dot{q}(\Dx)^2}{k}\right) +
\\
\T{m}{p}\left(\frac{1}
                   {\Dt}-2\frac{k}
             {\rho C_p (\Dx)^2}\right) = \frac{\T{m}{p+1}}
                                              {\Dt}
\end{array}
\end{equation}

We can define the dimensionless Fourier number $Fo=\frac{\alpha \Dt}{(\Dx)^2}$ and then express the preceding equation in term of this number~:

\begin{equation}
Fo
\left( \T{m-1}{p}+\T{m+1}{p}+\dot{q}\frac{(\Dx)^2}{k} \right) +
\T{m}{p} (1-2Fo) = \T{m}{p+1}
\end{equation}

The description for the chain extremities depends on the spatial initial and final conditions. In our case, we supposed that the sample is in a liquid at a temperature $T_\infty$ with a convection factor $h$. After defining the dimensionless Biot number $( Bi=\frac{h\Dx}{k} )$ we can write~:

\begin{equation}
\begin{array}{cc}
\T{N}{p+1}=2Fo 
\left(\T{N-1}{p}+Bi T_\infty + \dot{q}\frac{(\Dx)^2}
                                           {2 k}\right) +
\\
(1 - 2 Fo - 2 Bi Fo)\T{N}{p} 
\\
\mathrm{and} \\
\T{0}{p+1}=2Fo 
\left(\T{1}{p}+Bi T_\infty + \dot{q}\frac{(\Dx)^2}
                                         {2 k}\right) +
\\
(1 - 2 Fo - 2 Bi Fo)\T{0}{p}
\end{array}
\end{equation}

If we want this equation to have a physical meaning~: $(1 - 2 Fo - 2 Bi Fo) > 0$, we must impose a condition on \Dx and \Dt.

Since the discretization of space and time is not entirely free, we had to choose our steps carefully. The preceding inequality can be expressed in terms of \Dx and \Dt~:$$ \Delta t < \frac{(\Delta x)^2 \rho C_p}{2 (k+h\Delta x)} $$

Of course, this equation can always be fulfilled (by diminishing \Dt) but the corresponding increase in computing time can become unrealistic.

\noindent We implemented this algorithm in C++ (source avalaible upon request).

\begin{figure}[pt]
  \begin{center}
    \includegraphics[height=3 in,angle=270]{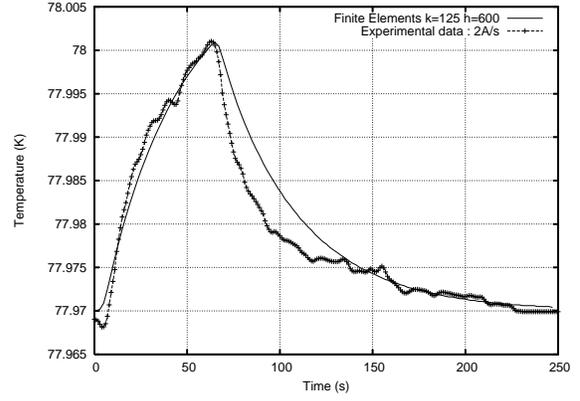}
    \caption{\footnotesize Fit with finite element analysis following the application of the 2A/s trapezoidal current pulse}
    \label{fig:fit:ef:1}
  \end{center}
\end{figure}

Fig~\ref{fig:fit:ef:1} represents the results of the finite element method for the power pulse with a 2A/s current slope with the experimental data. Again, similar results are obtained with the 0.2A/s pulse. The proposed modelization is accurate only for the increasing temperatures. This is an awaited result because our modelization is, for the moment, stricly unidimensional. As the preceding modelization shows us, there are heat leaks in our system and we must take them into account.

Indeed, with our finite element simulation, we take into account only one contribution~: the thermal conduction of heat along the sample. In order to take into account this phenomenon, we have to consider the two dimensions of the sample and introduce another parameter ($h^\prime$) that will describe the heat conduction along the y axis.

The simulations give us a thermal conductivity between 100 and 125 W.m\pui{-1}.K\pui{-1}. These results are consistent with the data cited in references \cite{thermique1} and \cite{goldacker2000} in which the thermal conductivity is seen to decrease with increasing gold concentration. The results for our 6\% gold sample falls between the 2.9\% and 11\% samples of ref~\cite{thermique1} and between the 4\% and 10\% samples of ref~\cite{goldacker2000}.

\section{\sc\bf Conclusion}

We have evaluated the thermal conductivity of \htsc with a very simple apparatus. Whereas the precision is not high (10\%), it is coherent with other thermal conductivity measurements made on BSCCO tapes. The use of three CERNOX sensors will gives us a better precision and will negate the influence of the thermal drift of the liquid nitrogen bath.

\end{document}